\documentclass[fleqn,10pt]{wlscirep}
\usepackage[utf8]{inputenc}
\usepackage[T1]{fontenc}

\usepackage{multirow}

\title{Role of non-linear data processing on speech recognition task in the framework of reservoir computing}

\author[1,*]{Flavio Abreu Araujo}
\author[2]{Mathieu Riou}
\author[3]{Jacob Torrejon}
\author[4]{Sumito Tsunegi}
\author[5]{Damien Querlioz}
\author[4]{Kay Yakushiji}
\author[4]{Akio Fukushima}
\author[4]{Hitoshi Kubota}
\author[4]{Shinji Yuasa}
\author[6]{Mark D. Stiles}
\author[2]{Julie Grollier}

\affil[1]{Institute of Condensed Matter and Nanosciences, Universit\'{e} catholique de Louvain, Place Croix du Sud 1, 1348 Louvain-la-Neuve, Belgium}

\affil[2]{Unité Mixte de Physique, CNRS, Thales, Université Paris-Sud, Université Paris-Saclay, 91767 Palaiseau, France.}

\affil[3]{Service de Physique de l'Etat Condensé, DSM/IRAMIS/SPEC CNRS UMR 3680 CEA Saclay, 91191 Gif-sur-Yvette Cedex, France}

\affil[4]{National Institute of Advanced Industrial Science and Technology (AIST), Spintronics Research Center, Tsukuba, Ibaraki 305-8568, Japan.}

\affil[5]{Centre de Nanosciences et de Nanotechnologies, CNRS, Université Paris-Sud, Université Paris-Saclay, 91405 Orsay, France}

\affil[6]{Physical Measurement Laboratory, National Institute of Standards and Technology, Gaithersburg, Maryland 20899-6202, USA.}

\affil[*]{flavio.abreuaraujo@uclouvain.be}

\usepackage{interval}
\newcommand{\Didx}{\sigma}
\newcommand{\Xm}{\mathbf{X}_\Didx}
\newcommand{\Zm}{\mathbf{Z}_\Didx}
\newcommand{\Rm}{\mathbf{R}_\Didx}

\newcommand{\NL}{\alpha}
\newcommand{\idx}{{f\tau,\Didx}}
\newcommand{\Nf}{N_f} % Number of channels
\newcommand{\Ntau}{N_\tau} % Number of time steps
\newcommand{\Ntheta}{N_\theta} % Number of nodes (neurons)
\usepackage{soul}

\newcommand{\SpectroHP}{Spectro HP}
\newcommand{\onlinecite}[1]{\hspace{-1 ex} \nocite{#1}\citenum{#1}}
\begin{abstract}
The reservoir computing neural network architecture is widely used to test hardware systems for neuromorphic computing. One of the preferred tasks for bench-marking such devices is automatic speech recognition. This task requires acoustic transformations from sound waveforms with varying amplitudes to frequency domain maps that can be seen as feature extraction techniques. Depending on the conversion method, these transformations sometimes obscure the contribution of the neuromorphic hardware to the overall speech recognition performance. Here, we quantify and separate the contributions of the acoustic transformations and the neuromorphic hardware to the speech recognition success rate. We show that the non-linearity in the acoustic transformation plays a critical role in feature extraction. We compute the gain in word success rate provided by a reservoir computing device compared to the acoustic transformation only, and show that it is an appropriate bench-mark for comparing different hardware. Finally, we experimentally and numerically quantify the impact of the different acoustic transformations for neuromorphic hardware based on magnetic nano-oscillators.
\end{abstract}

\begin{document}

\flushbottom
\maketitle

\thispagestyle{empty}

%\noindent Please note: Abbreviations should be introduced at the first mention in the main text – no abbreviations lists. Suggested structure of main text (not enforced) is provided below.

\section*{Introduction}

Artificial neural network algorithms outperform humans on recognition tasks like image or speech recognition, by leveraging deep networks of interconnected non-linear units called formal neurons \cite{LeCun2015}. The goal of neural networks is to extract the features and classify input data through learned non-linear transformation. Running such algorithms on a classical computer is costly energetically: to overcome this issue, neuromorphic approaches \cite{Mead1990, Mead2012} propose to implement them physically. In particular, reservoir computing \cite{Maass2002, Jaeger2004} is a kind of recurrent neural network that has been widely used to test the efficiency of hardware for neuromorphic computing \cite{Appeltant2011, Paquot2012, Torrejon2017} because it has a simplified architecture and learning procedure. The input is sent to a neural network with fixed recurrent connections called a reservoir. The goal of the reservoir is to separate the different kinds of inputs, such that after this transformation, the classification can be done by a linear transformation. The response of the neurons of the reservoir are combined linearly with trained connections to construct the output. Since the connections in the reservoir are random and fixed, it is easier to fabricate it in hardware and then train the output connections, often emulated in software, with linear regression.

\begin{figure}[ht]
    \centering
    \includegraphics[scale=1.25]{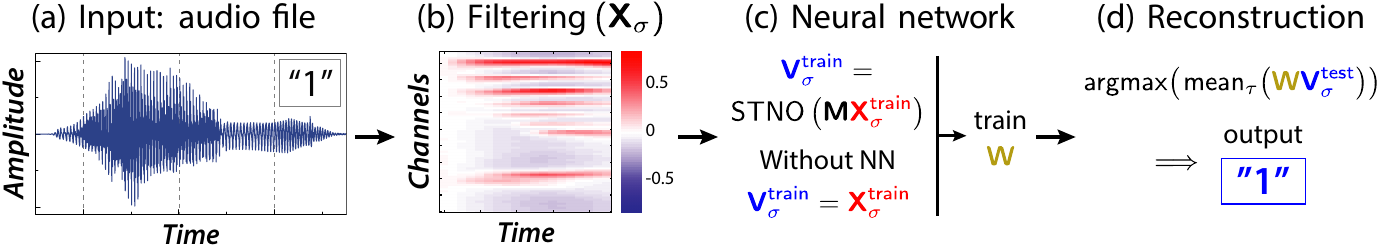}
    \caption{{\bf Principle of spoken digit recognition.} (a) Audio waveform corresponding to the digit 1 pronounced by speaker 1. (b) Filtering to frequency channels for acoustic feature extraction. The signal during each time interval $\tau$ is decomposed in $\Nf$ frequency channels. The cochlear model filters each point of the audio waveform in 78 frequency channels (13 in the case of the MFCC model and 65 for the spectrogram model). The frequency channels are concatenated in intervals of duration $\tau$ to form the filtered input. (c) The filtered input is injected in the neural network or directly used to construct the output (No neural network). The neural network is composed of $N$ interconnected filtered inputs. (d) For each digit, the response of the neural network (or directly the filtered output) is constructed from a linear combination of neuron states $V_{\theta\tau,\sigma}$ (there are 10 classifiers in total).}
\label{fig1}
\end{figure}

Speech recognition is a widely used class of benchmark tasks performed to test the efficiency of a neural network. It is especially employed in the case of reservoir computing because the recurrent connections of the reservoir create an intrinsic memory that is useful to classify time-varying inputs. Generally, this task requires frequency decomposition \cite{Davis1980, Verstraeten2005, Hinton2012} prior to the neural network because the acoustic features are contained in the frequency rather than in the amplitude of the time-varying signal. These decompositions return the amplitude of the signal in different frequency channels as a function of time. The neural network then extracts the acoustic features contained in the frequency information. Several frequency decomposition methods have been reported in the literature: Mel-frequency cepstral coefficients (MFCC) and Lyon’s cochlear model (cochleagram) are the most common methods since they mimic the filtering that occurs biologically \cite{Davis1980, Lyon1982, Slaney1988}. However, the actual contribution of the acoustic filter to the total speech recognition rate is generally not investigated while performing speech recognition benchmarks with reservoir computing hardware, even if its influence on the final recognition rate may not be negligible \cite{Torrejon2017}. Furthermore, both of these methods were developed before reservoir computing became popular and, thus they were designed to extract the useful features of an audio signal independent of modern machine learning.

Here, we first show how the choice of different filtering methods drastically affects the final speech recognition rate. We quantify the contributions of the acoustic filtering and the neural network, respectively, for a spoken digit recognition task using four frequency decomposition methods with different non-linear characters: Lyon's ear cochleagram, MFCC filter, linear spectrogram ($\Re{(\text{Spectro})}$), and \SpectroHP~(\SpectroHP~$=| \sin{\sqrt{|\Re{(\text{Spectro})}|}} | - | \cos{\sqrt{|\Im{(\text{Spectro})}|}} |$). In a first step, we show that the cochleagram, the \SpectroHP and the MFCC filter are powerful stand-alone features extractors that can achieve by themselves (without additional processing by a neural network) very high recognition levels: up to 95.8~\%, 89.0~\%, and 77.2~\% for cochleagram, \SpectroHP, and MFCC, respectively.  In contrast, the linear spectrogram never achieves recognition levels statistically better than random sampling, 10~\%.  However, by adding various levels of non-linearity to the real part of the spectrogram we were able to show a large increase of the recognition rate from about 10~\% (linear) to 88~\% (strong non-linearity). These results indicate that the high recognition level of the cochleagram and MFCC approaches is mainly due to the non-linear character of these frequency decomposition methods and not to the reservoir itself. 

In a second step, we evaluate the gain in recognition rate provided by a particular hardware approach to reservoir computing, based on magnetic nano-oscillators. In order to compare to other hardware implementations in the literature, we model a neural network based on a single  dynamical non-linear magnetic node in the framework of the reservoir computing approach \cite{Appeltant2011, Paquot2012, Brunner2013, Vandoorne2014, Torrejon2017, Riou2017_IEDM_Paper}. We find that the contribution of the neural network is dominant for linear spectrogram filter and only plays a small role for the non-linear cochleagram and MFCC filter. Finally, we present experimental results using a non-linear and tunable magnetic nano-oscillator exhibiting excellent agreement with our simulations.

\begin{figure}[!hb]
    \centering
    \includegraphics[scale=1]{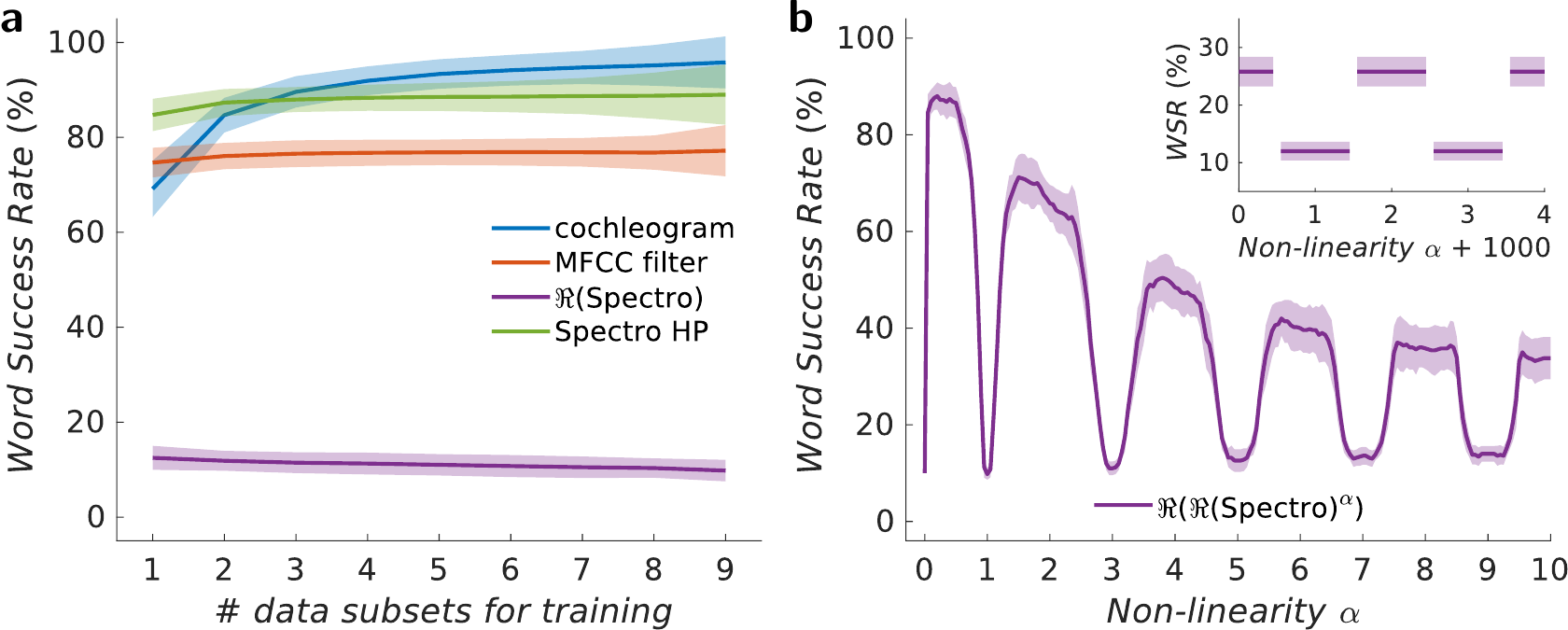}
    \caption{{\bf Spoken digit recognition for filtered inputs.} (a) Spoken digit cross-validated test recognition rates as a function of the number of data subsets $N$ used for training (total size of the training set $5\times 10 \times N$) of the filtered input (without neural network) corresponding to four different methods: cochleagram, MFCC filter, \SpectroHP and linear spectrogram ($\alpha = 1$). (b) Spoken digit recognition as a function of non-linear coefficient for spectrogram methods (Inset: Word success rate for large non-linear coefficient values from 1000 to 1004). Here, 9 data subsets (90~\% of the database) are used for training our reservoir computing model and the remaining subset (10~\% of the database) is used to perform the recognition task. The shaded region corresponds to the uncertainty of the recognition rate, here the standard deviation).}
\label{fig2}
\end{figure}

\section*{Methods}

%Topical subheadings are allowed. Authors must ensure that their Methods section includes adequate experimental and characterisation data necessary for others in the field to reproduce their work.

We perform a benchmark task called spoken digit recognition that is common in the reservoir computing community for software \cite{Verstraeten2005} and hardware \cite{Paquot2012, Larger2012, Brunner2013, Dejonckheere2014, Vinckier2015, Torrejon2017, Brunner2018, Penkovsky2018} implementations. The input data, taken from the TI-46 database, are audio waveforms of clean isolated spoken digits (0 to 9) pronounced by five different female speakers (see example in Fig.~\ref{fig1}a), as it is usual in the hardware reservoir computing community\cite{Verstraeten2005, Appeltant2011, Larger2012, Paquot2012, Duport2012, Martinenghi2012, Brunner2013, Martinenghi2014, Torrejon2017, Larger2017, Brunner2018}.

The chosen part of the TI-46 spoken digit database contains 500 (5 speakers $\times$ 10 digits $\times$ 10 utterances) audio files, which we index using the Greek letter $\sigma$.
To perform speech recognition on these spoken digits, each audio temporal trace in the database is transformed from time-domain to a mixed time/frequency domain with different acoustic filters, two of which are known to create a better representation of human voice characteristics. These acoustic filters give rise to different instances of our speech database containing the following elements: $X_{f\tau,\sigma}^\text{filter}$ where filter $\in \{\text{Cochlear}, \text{ MFCC}, \text{ Spectro}, \text{ \SpectroHP}\}$, $f$ is the index for the different frequency channels, and $\tau$ is the index of a new time representation that depends on the time frame window used while performing the time- to frequency-domain transformation. 
The number of time steps $N_\tau$ naturally depends on the digit length, while the number of frequency channels $N_f$ only depends the type of acoustic filter. For instance, $N_f^\text{Cochlear} = 78$, $N_f^\text{MFCC} = 13$, and $N_f^\text{Spectro} = 65$ while $N_\tau^\text{Cochlear}$ ranges from 16 to 41, $N_\tau^\text{MFCC}$ ranges from 31 to 83, and $N_\tau^\text{Spectro}$ ranges from 24 to 67.
Digits with $N_\tau$ lesser than the maximum value are padded with zeros.

The construction of the supervised learning task in the reservoir computing framework starts with associating each digit $\mathbf{X}_\sigma$ with its corresponding target $\mathbf{T}_\sigma \in \mathbb{R}^{N_d \times N_\tau}$ where $N_d$ is the number of categories to classify (here $N_d = 10$ as the goal is to recognise the 10 different digits). Each target matrix $\mathbf{T}_\sigma$ is constructed column-wise and the $N_\tau$ columns correspond to the same target vector $\mathbf{t}_\sigma$ (the target vector $t_{d,\sigma}$ with $d\in[0..9]$ is zero almost everywhere but is one where $d$ is equal to the corresponding digit number). The $\mathbf{T}_\sigma$ matrices would allow us to perform $\tau$-wise recognition (partial digits for instance) but in this study we choose to make entire-digit-wise recognition by averaging out the estimated target matrices $\hat{\mathbf{T}}_\sigma$ over the different columns ($\tau$ direction) to end up with estimators $\hat{\mathbf{t}}_\sigma$ of the target vectors $\mathbf{t}_\sigma$ as shown below (see Eq. (\ref{Eq:test})).

Our reservoir is a time multiplexed single device as described in Ref.~\onlinecite{Torrejon2017}.  Rather than a set of $N_\theta$ physical neurons, our reservoir consists of a single physical neuron evaluated at $N_\theta$ periodic times.  To input the data to these virtual neurons, we multiply each value by a time series of length $N_\theta$ consisting of ones and minus ones and send the resulting time series to the single device.  The output of the reservoir is determined by the resulting state of the device at each of the $N_\theta$ times for each element of the input data string.  This output is multiplied by the output weight matrix to give the results.  Training consists of determining the optimum set of output weights, which can be found through straightforward linear algebra.

The key computational concept supporting the reservoir computing approach is a nonlinear dynamical transformation of the processed information, i.e. sending the input data to a new space, in which simple linear algebra gives the read-out of the results\cite{Lukosevicius2009}. In this work, the non-linear transformation is the purpose of our spin-torque nano oscillator represented by the function $\text{STNO}(\cdot)$ in Eq. (\ref{Eq:RC}). The information is then encoded and injected into this nonlinear dynamical system after flattening the data and multiplying each element of the flattened $\mathbf{X}_\sigma$  by a random binary mask $\mathbf{M} \in \mathbb{R}^{N_\theta \times N_f}$ of 1's and -1's.  This binary masks starts the time-multiplexing technique as the value times the mask gives the input to each virtual neuron. As a result, the mask distributes the frequency content of each time step $\tau$ of the input data into a fixed neural network layer (the reservoir) of $N_\theta$ nodes. To summarise, Eq. (\ref{Eq:RC}) shows the details of our Reservoir Computing implementation:
\begin{equation}
\mathbf{X}_\sigma \rightarrow \mathbf{V}_\sigma: \begin{cases}
\mathbf{x}_\sigma = \text{flatten}\left(\mathbf{M} \mathbf{X}_\sigma\right) & x_{i,\sigma} = \text{flatten}\left(\sum_f M_{\theta f} X_{f\tau,\sigma}\right) \text{ with } i \in [1..N_\theta N_\tau],\\
\mathbf{v}_\sigma = \text{STNO}(\mathbf{x}_\sigma) & v_{i,\sigma} = \text{STNO}(x_{i,\sigma}),\\
\mathbf{V}_\sigma = \text{reshape}(\mathbf{v}_\sigma) & V_{\theta\tau,\sigma} = \text{reshape}(v_{i,\sigma}),
\end{cases}
\label{Eq:RC}
\end{equation}
where flatten$(\cdot)$ takes a $m$ by $n$ matrix as input and outputs a vector of length $mn$ and reshape$(\cdot)$ does the reverse operation, i.e. takes a vector of length $mn$ as input and outputs a $m$ by $n$ matrix.

Training (learning) is performed using a simple linear classifier. In this work, good performance is achieved using the Moore-Penrose pseudo-inverse after building the weight matrix $\mathbf{W}$ optimisation problem with a subset of $N_\text{train}$ digits $\mathbf{W} \left[ \mathbf{V}_1, \mathbf{V}_2, ..., \mathbf{V}_{N_\text{train}} \right] = \left[ \mathbf{T}_1, \mathbf{T}_2, ..., \mathbf{T}_{N_\text{train}} \right]$:
\begin{equation}
    \mathbf{W}  = \left[ \mathbf{T}_1, \mathbf{T}_2, ..., \mathbf{T}_{N_\text{train}} \right] \left[ \mathbf{V}_1, \mathbf{V}_2, ..., \mathbf{V}_{N_\text{train}} \right]^{-1} .
\end{equation}
No regularisation technique is used. The testing (recognition) step is then achieved using the computed weights $\mathbf{W}$ applied to the complementary (unseen) subset of $N_\text{test}$ digits:
\begin{equation}
\begin{matrix}
\hat{\mathbf{T}}_\sigma = \mathbf{W} \mathbf{V}_\sigma & \hat{T}_{d\tau,\sigma} = \sum_{\theta} W_{d\theta} V_{\theta\tau,\sigma}\\
\displaystyle \hat{\mathbf{t}}_\sigma = \text{mean}_\tau\left(\hat{\mathbf{T}}_\sigma\right) &  \displaystyle \hat{t}_{d,\sigma} = \frac{1}{N_\tau}\sum_{\tau=1}^{N_\tau} \hat{T}_{d\tau,\sigma}
\end{matrix} .
\label{Eq:test}
\end{equation}
The estimator for a specific digit is given by $\hat{d}_\sigma = \text{argmax}(\hat{\mathbf{t}}_\sigma)$ (this corresponds to the Winner-Takes-All strategy, adequate for the present task). Digit $\sigma$ is well recognised when $\hat{d}_\sigma = \text{argmax}(\mathbf{t}_\sigma)$ and the main performance estimator used in this work is the Word Success Rate (WSR) and  corresponds to the percentage of well recognised digits over the total number of digits to recognise ($N_\text{test}$). Another common performance estimator, useful for identifying overfitting issues, is the Mean Squared Error (MSE): MSE($\hat{\mathbf{t}}_\sigma$) = $\mathbb{E}\left[(\hat{\mathbf{t}}_\sigma - \mathbf{t}_\sigma)^2\right]$.

In all cases, the training and testing sets do not overlap. To avoid any learning bias while selecting samples randomly from the database when choosing the training and testing sets, we organise the 500 input files into 10 subsets of 50 files.  Each subset contains one utterance of each digit pronounced by each of the five speakers.  A random selection would produce an over-representation of some speakers and an under-representation of others. We take $N$ utterance subsets (50 audio files, one for each digit and each speaker) for training (total training set size $N\times 50$), and $10-N$ utterance subsets for testing (total testing set size $(10-N)\times 50$).  To minimise the fluctuations that occur in the results due to random choices between the training and testing sets, we employ a cross-validation technique and therefore average over all possible choices.  That is, when $N$ utterance subsets are used for training, we average over $10!/[N!(10-N)!]$ possible ways to choose the training and testing sets.  This procedure also allows us to determine a width to the distribution of individual outcomes indicated by the shaded regions in Fig.~\ref{fig2}. All word success rate results reported in the paper are cross-validated test results.

In reservoir computing, training is fast and always converges due to the basic linear algebra algorithms.  This behaviour stands in contrast to standard  recurrent neural-network approaches for which learning can be time consuming and does not necessarily converge to the desired solution. In reservoir computing, the learning process only modifies the read-out weights whereas in other types of recurrent neural network it modifies the weights in all the other constituent layers in the neural network under complex feed-forward/back-propagation algorithms.

The contribution of the frequency filtering and the reservoir computing, respectively, are then analysed separately. In order to evaluate the impact of the frequency filtering on the input separation capability, a linear classifier is trained directly on the different frequency channels. The classification results with both influence of the frequency filtering and the reservoir are computed by injecting the filtered input in a neural network composed of $\Ntheta$ interconnected neurons. Here, we use $\Ntheta = 400$ input neurons that are connected to all of the frequency channels for each time interval $\tau$, Fig.~\ref{fig1}c, as this number allowed reaching maximum test accuracy. In the framework of reservoir computing, these fixed connections have random weights. To reach high classification rate, 400 neurons are sufficient \cite{Appeltant2011, Torrejon2017}. The features of the magnetic neurons that we consider are specified in section IV. A linear classifier is trained to map the neuron outputs to the desired results. The contribution of the reservoir to the ultimate success is extracted from the results by subtracting the success rate found using only the frequency filtering methods.

\section*{Acoustic filter: role of non-linearity}

First, we compute the digit recognition rate as a function of the number of utterances used in training for the cochleagram and the MFCC methods as shown in Fig.~\ref{fig2}(a). The recognition rate increases with the number of trained utterances and then saturates in the case of the cochlear model. It remains almost constant for the MFCC model. Both filters achieve a high recognition rate. In particular, the cochlear model is an excellent acoustic feature extractor with recognition rates up to 95.8~\% (for 9 trained utterances) whereas the MFCC filter is less powerful, reaching recognition rates up to 77.2~\%.

These filters are commonly used for speech recognition tasks, because of their similarity to audio signal processing in biological ears, which perform complex frequency decompositions with high non-linearities. Both MFCC and cochlear methods use non-linearities to transform the audio data. For MFCC the transformed representation corresponds to the log-energy of the Mel frequency filter output \cite{Davis1980}. In the cochleagram approach, the main non-linear ingredient corresponds to a set of interconnected automatic gain controls \cite{Lyon1982, Slaney1988}. The successful separation of the data achieved by these filtering methods appears to be mainly due to the non-linear character of the transformation with a moderate influence of the kind of non-linearity (similarly to reservoirs that can have different kinds of non-linearity that work).

To establish the critical role of the non-linearity contained in the filtering methods to recognition performance, we start by investigating the separation achieved by a very simple linear spectrogram filter. This filter is based on standard Fourier transforms of the audio input over finite time windows. The Fourier transform is a linear operation that outputs a real and an imaginary part. We consider only the real part in the following in order to avoid introducing non-linearities by computing the norm. After the Fourier transforms, $\Zm$ is the matrix of the real parts of the spectrogram with dimension $\Nf \times \Ntau$ where $\Nf$ is the number of frequency channels and $\Ntau$ is the number of time steps, which depends on the particular digit. We normalise the data, $\Xm = \Zm/\max(|\Zm|)$ and $X_\idx \in [-1, 1]$ for $f \in \{1..\Nf\}$ and $\tau \in \{1..\Ntau\}$. The normalisation is crucial to ensure that there exists at least one $X_\idx$ that is equal either to 1 or to -1 for each $\Xm$ when non-linearities are introduced into the transform.

To study the influence of a non-linear transformation on the normalised input data $\Xm$, we choose to apply a point-wise operation, namely the exponent $\NL \in \mathbb{R}$, giving rise to the transformation on each element of $X_{f\tau,\sigma}^\text{filter} \rightarrow (X_{f\tau,\sigma}^\text{filter})^{\NL}$. The impact of the non-linear exponent $\NL$ on the recognition rate is shown in Figure~2(b). The recognition rate oscillates strongly as a function of the non-linear exponent and decreases for large $\NL$. Some particular values of the recognition rate can be easily understood. For $\NL = 0$: $\forall$ $i$ and $j$, $X_\idx = 1$, and it becomes impossible to discriminate between different digits $\Xm$ and the success rate is equal to $10$~\% (random choice). As $\NL$ approaches zero, the success rate decreases drastically and drops to 10~\%. For such exponents, all inputs get mapped to the same output making data separation impossible. For $\NL = 1$ the real part of the spectrogram corresponds to a linear transformation of the input data, thus there is no non-linear data separation and the word recognition rate $\simeq 10$~\% (random choice).

The evolution shown in Figure~2(b) can be understood by decomposing the exponent $\NL$ into an integer part $n \in \mathbb{N}$ and a real part $\varepsilon \in \mathbb{R}$ around $n$ ($\varepsilon \in~]-0.5, 0.5]$): $\NL = n + \epsilon$.  
For $X_\idx < 0$, $X_\idx^- \rightarrow (X_\idx^-)^{n+\epsilon} = |X_\idx^-|^{n+\epsilon}(-1)^n (\cos(\pi\epsilon) + i\sin(\pi\epsilon))$ and for $X_\idx \geqslant 0$, $X_\idx^+ \rightarrow (X_\idx^+)^{n+\epsilon} = |X_\idx^+|^{n+\epsilon}$. For simplicity, we choose to consider only the real part of the data obtained after applying the non-linearity, so $\Rm = X_\idx \rightarrow \Re(X_\idx^{n+\epsilon})$:\\
\begin{equation}
	\Rm = \left\{\begin{aligned}
	\Re((X_\idx^-)^{n+\epsilon}) &= |X_\idx^-|^{n+\epsilon}(-1)^n \cos(\pi\epsilon)\\
	\Re((X_\idx^+)^{n+\epsilon}) &= |X_\idx^+|^{n+\epsilon}.\\
	\end{aligned}\right.
\label{eq:NL}
\end{equation}

From Eq. (\ref{eq:NL}), for $X_\idx < 0$ there is an additional factor $(-1)^n \cos(\pi\epsilon)$ compared to $X_\idx > 0$. Consider the particular case where $\epsilon = 0$, then in the case of values of $X_\idx$ that were initially negative, the values $R_\idx$ have the sign $(-1)^{n}$. So, depending on the parity of $n$, there are two  possibilities. If $n$ is even, there is at least one value $R_\idx$ in $\Rm$ equal to 1 for each $\Rm$ (digit in the database). If $n$ is odd, the $\Rm$ digits originating from an input $\Xm$ where at least one $X_\idx = -1$ have a corresponding $R_\idx = -1$ (at least one $R_\idx = 1$ otherwise). Therefore, the oscillating behaviour of the success rate shown in Fig.~\ref{fig2}b is related to what happens to the negative input data as shown in Eq. (\ref{eq:NL}).

The poorer performance for the recognition task for odd $n$ comes from the fact that the phase from the Fourier transform is essentially arbitrary.  When $n$ is even, the important elements of $\Rm$ are always positive, but for odd $n$ they are sometimes positive and sometimes negative.  The greater variation in the latter case makes it essentially impossible for the neural network to connect the input date to the appropriate output.  This behaviour is most easily seen in the limit that $n$ becomes large as shown in the inset of Fig.~\ref{fig2}b.

From Eq. (\ref{eq:NL}) we can evaluate the effect of our non-linear transformation for $n \rightarrow \infty$:
$\displaystyle\lim_{n \rightarrow \infty} |X_\idx|^n = 0$ for $|X_\idx| < 1$ and $\forall n$ $|X_\idx|^n = 1$ when $|X_\idx| = 1$.
In practice, due to the numerical truncation on a computer, for $n \gg 100$ and $|X_\idx| < 1$, $|X_\idx|^n = 0$. So, for very large $n$, the resulting vector $\Rm$ contains only zeros and at least one element that is equal to 1 or -1 after the non-linear transformation.

There are 500 digits in our spoken digit database and for very large odd values of $n$, there are 253 $\Rm$ vectors with one $R_\idx = 1$ and 247 with one $R_\idx = -1$. For each of these vectors, all other elements are mapped to zero. For large even values of $n$, all the 500 $\Rm$ contain one $R_\idx = 1$ with all others equal to zero.  In this large-exponent limit, the classification task is simple to understand. The non-linearity selects the largest magnitude frequency/time component from the transformed audio file. If this component is constant between speakers, the digit can be identified. For large values of $n$, there are 2 different success rate values depending on the parity of $n$. As shown in the inset of Fig.~\ref{fig2}b, for large values of $\NL$ ($\NL > 1000$), the success rate behaviour tends to a square function alternating between very low values (12~\%) around odd values of $\NL$, i.e. $\NL \in \interval[open]{2n+0.5}{2n+1.5}$, and a slightly higher value (25.8~\%) around even values of $\NL$, i.e. for $\NL \in \interval[open]{2n-0.5}{2n+0.5}$, where $n \in \mathbb{N}$ (for large values of $\NL$, when $\NL = n+0.5$, the success rate is not defined). The difference arises because of the random phases that arise from the Fourier transform. For even $n$, the phases are irrelevant, but for odd $n$, sometimes the value gets mapped to one and sometimes to minus one, making it much more difficult to classify.

Overall, as shown in Figure~2(b), for a wide range of values of $\NL$, the non-linearity drastically improves the recognition rate. In particular, the recognition rate is very high for low exponents $\NL$. An optimum non-linearity is reached for $\NL = 0.2$ providing the highest recognition rate of 88~\%, which is comparable to those obtained for the cochleagram. 

\begin{figure}[ht]
    \centering
	\includegraphics[scale=.8]{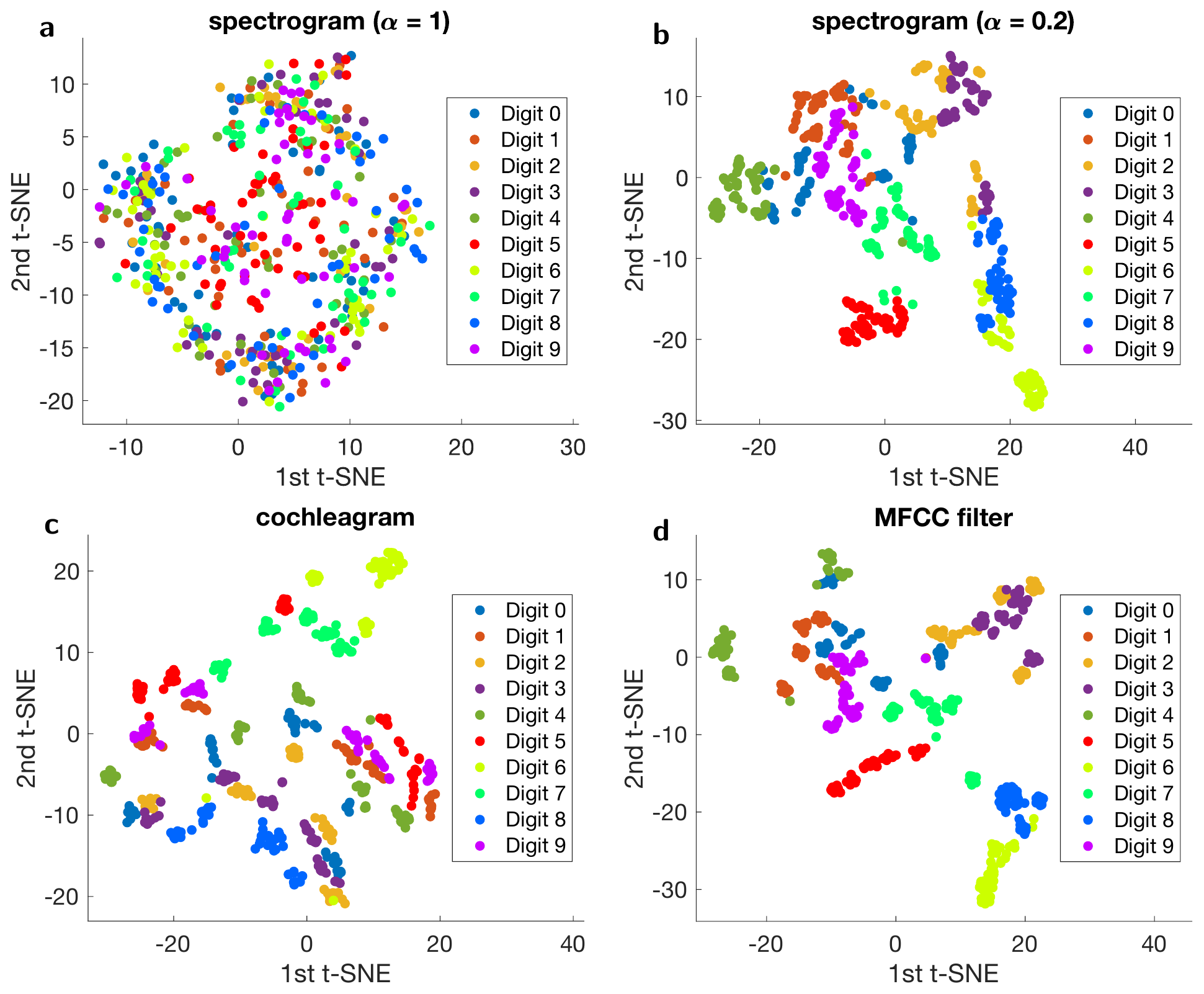}
	\caption{2D representation of the two $t$-SNE components for: (a), the spectrogram with $\NL = 1$, (b), the spectrogram with $\NL = 0.2$, (c), the cochleagram, and (d), the MFCC filtering methods.}
	\label{fig3}
\end{figure}

We use a $t$-Distributed Stochastic Neighbour Embedding ($t$-SNE) technique \cite{Maaten2008} to represent our $N_f$ channels data in a 2D plot (see Fig.~\ref{fig3}) in order to visualise how the data separation occurs and understand the recognition capacity of the different filtering methods. $t$-SNE is a nonlinear dimensionality reduction technique used for embedding high-dimensional data into a low-dimensional space of two or three dimensions. During the data reduction, the probability of two vectors to be neighbours is conserved, allowing visualisation of the structure in the data. Each digit is represented by coloured dots for all data points of the utterances. For instance, Fig.~\ref{fig3}a shows that for the spectrogram with $\NL = 1$ (linear) for which the recognition rate is about 10~\% (random choice), there is no data separation as all the coloured points seem to be randomly distributed. In particular the digits of a same class do not form separated clusters. On the other hand, for the spectrogram with $\NL = 0.2$ (optimal non-linearity), data separation can clearly be seen in Fig.~\ref{fig3}b correlating with the better recognition rate of 88~\% compared to the linear spectrogram ($\alpha = 1$). Furthermore, $t$-SNE shows data separation capability for both the cochleagram and the MFCC filter. As shown in Fig.~\ref{fig3}c and \ref{fig3}d, well defined clusters corresponding to the spoken digits appear and corroborate the high recognition rates exhibited by the two filtering methods.

To summarise this section, we show that the non-linear transformation applied to the input data by the MFCC filter and the cochleagram plays a similar role as the non-linear nodes in the reservoir neural network prior to the linear classifier. We highlight that these stand-alone feature extractors  perform data separation due to their internal non-linear transformations.  We indeed obtain recognition performance that are close to what is found with these approaches by adding a simple non-linear transformation to the individual elements of the conventional spectrogram. Depending on the non-linearity, the recognition rate can strongly vary from around 10~\% to 95.8~\%.

\section*{Neural network: Reservoir computing based on non-linear oscillators}

Having shown that non-linear filtering methods can by themselves achieve high recognition rates, we turn to evaluating the gain in overall performance provided by a reservoir neural network taking as inputs the output of these acoustic filters. We implement the reservoir with a single non-linear oscillator \cite{Appeltant2011}. In this approach,
recurrent chains of non-linear transformations occur in time instead of space. The loss of parallelism is compensated by time-multiplexing, in turn requiring that the input be preprocessed. To do that, each point of interval $\tau$ in Fig.~\ref{fig1}(b) is multiplied by a random binary matrix (of dimensions $\Nf \times \Ntheta$) to induce transient behaviour. This transformation is linear and does not affect the final recognition rate. Each point of the input audio file is converted in a binary sequence of duration $\tau$ composed of $\Ntheta = 400$  points separated by time steps $\theta$. The time step $\theta$ is set shorter than the relaxation time of the oscillator to keep the oscillator in the transient regime and generate temporal cascades at each sequence $\tau$ of the pre-processed input.

We have developed a simple model based on a non-linear magnetic oscillator \cite{Slavin2009} taking into account the main ingredients for neuromorphic computing: non-linearity (square root dependence of the amplitude on the input current) and memory (relaxation time of the oscillator between two different output voltage levels). The dynamics of the evolution of the oscillator output microwave voltage $v^\text{osc}_i$ as a function of the input voltage $v^\text{in}_i$ at time step $i$ can be solved numerically \cite{Riou2017_IEDM_Paper}:
\begin{equation}
\label{eq:Vosc}
    v^\text{osc}_i = v^\infty_i \left( 1 - e^{-\Delta t / T_\text{relax}} \right) + v^\text{osc}_{i-1} \cdot e^{-\Delta t / T_\text{relax}},
\end{equation}%
where $T_\text{relax}$ is the relaxation time towards the asymptotic value $v^\infty_i$ given by \cite{Grimaldi2014}:
\begin{equation}
v^\infty_i = c \sqrt{I_\text{DC} - v^\text{in}_i / R - I_\text{c}},
\end{equation}
with $c$, a constant related to the initial bias condition, i.e. the initial emitted voltage of oscillator, $R$ the DC resistance of the oscillator and $I_\text{c}$ the threshold current above which auto-oscillations can occur. In order to simulate the oscillator response to a time varying input ($V^\text{in}_i$), we solve Eq.~(\ref{eq:Vosc}) numerically with the following parameters: $\Delta t = 5$~ns, $V^\text{in}_i / R = \pm 3$~mA, $I_\text{DC} = 6$~mA, $I_\text{c} = 4.9$~mA, $T_\text{relax} = 410$~ns.
These parameters, which constitute huperparameters of our system, are extracted from experiments as reported elsewhere \cite{Torrejon2017}.

\begin{figure}[ht]
    \centering
	\includegraphics[scale=1]{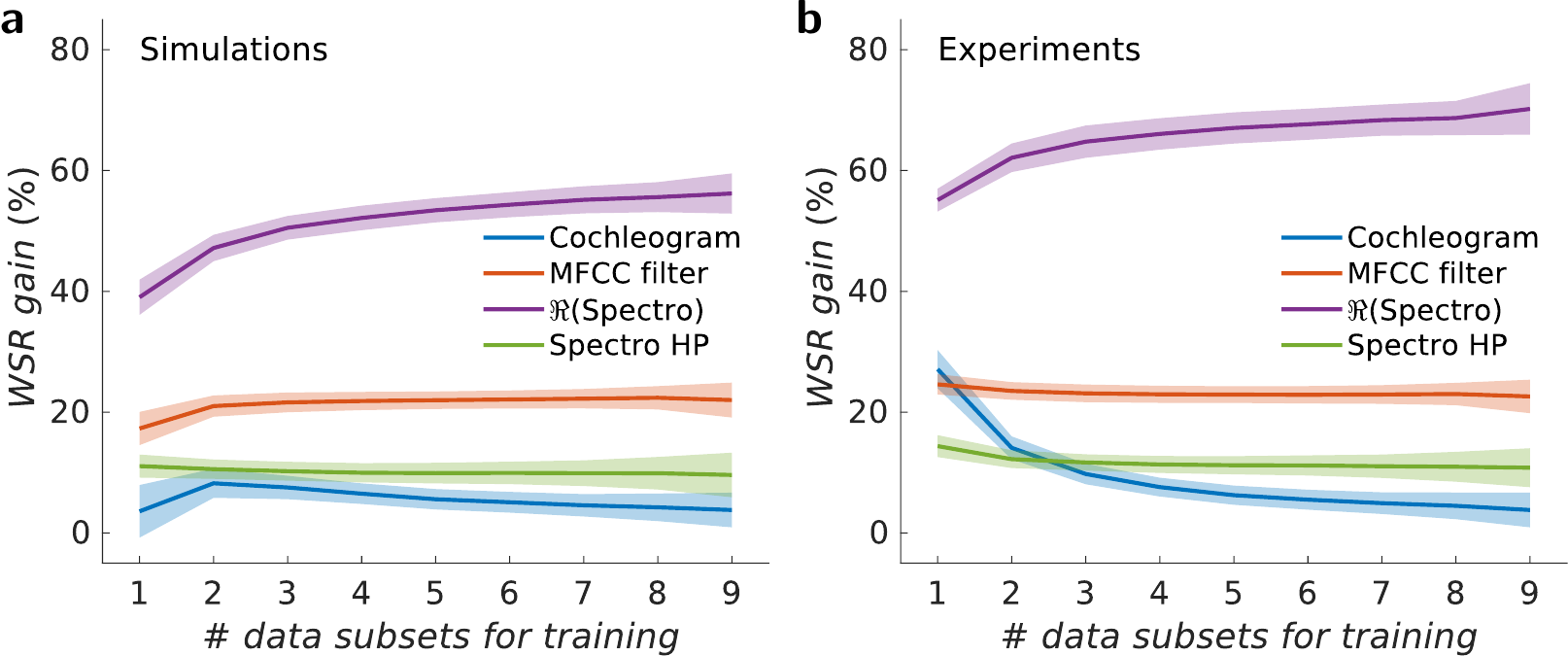}
	\caption{{\bf Spoken digit recognition for a neural network.} (a) Spoken digit gain on cross-validated test recognition rates as a function of the number of subsets $N$ used during training for a non-linear oscillator modelled with Eqs.(2-3) and (b) experimental spin torque nano-oscillator driven by spin polarised current. The coloured region corresponds to the uncertainty of the recognition rate, here twice the standard deviation.}
	\label{fig4}
\end{figure}

Even if the recognition rate by the non-linear filters (MFCC and cochleagram) is already high, there is still room for improvement with the inclusion of a recurrent neural network. The increase in the recognition rate induced by the emulated non-linear oscillator is shown in Fig.~\ref{fig4}a. We determine the increase in recognition rate due to the neural network by subtracting from the total recognition rate the contribution from acoustic filters previously calculated in Fig.~\ref{fig2}(a) and normalising the result with the total recognition rate. The gain provided by the non-linear oscillator is low for the non-linear filters (training over 9 data subsets): 3.8~\% for the cochlear method, 9.6~\% for the {\SpectroHP} method, and 22~\% for MFCC method. The increase is small because the total recognition rate (filter + network) is close to a perfect success rate: up to 99.6~\%, 98.6~\%, and 99.2~\% for the cochleagram, \SpectroHP, and the MFCC filter respectively. On the other hand, the neural network drastically improves  the recognition gain up to 55.6~\% (for 9 data subsets trained) but the final recognition rate (filter + neural network) is not as good, around 65.2~\%.
The simulations have been obtained for a specific neural network based on the non-linear dynamics of an oscillator with time multiplexing in the framework of reservoir computing. As mentioned earlier, we choose this particular framework because it is frequently used for hardware implementations. However, these conclusions hold for very general types of (spatial or temporal) neural networks and learning processes because the limitation of the gain in recognition in the case of MFCC, \SpectroHP, and cochlear filters is not due to the neural network but to the already excellent separation properties of the filtering.

We compare these simulations to the behaviour of an experimental non-linear oscillator. In particular, we choose a magnetic nano-oscillator that was recently demonstrated to be an excellent building block for neuromorphic computing \cite{Torrejon2017, Riou2017_IEDM_Paper, Romera2018}. This kind of oscillator is small (nanoscale), performs low power computing, has a high signal to noise ratio for high reliable computation, and  allows a tunable non-linearity through the spin transfer torque mechanism. Our nanoscale oscillators are circular magnetic tunnel junctions, with a 6~nm thick FeB free layer and a diameter of 375~nm. The magnetisation in the FeB layer has a vortex structure as its ground state for these dimensions. In a small region called the vortex core, the elsewhere in-plane-curling magnetisation points out of the plane. Under dc current injection, the core of the vortex steadily gyrates around the centre of the dot with a frequency in the range 250~MHz to 400~MHz. Vortex dynamics driven by spin-torque are well-understood, well-controlled and have been shown to be particularly stable (more details can be found elsewhere \cite{Tsunegi2014}).

The experimental implementation of the spoken digit recognition task is described in Ref. \cite{Torrejon2017}. The preprocessed input signal (filtered digits with time multiplexing) is generated and sent to the sample using an arbitrary waveform generator. Then, the microwave voltage across the magnetic tunnel junction is measured by a real time oscilloscope and fast oscillations are observed. The amplitude of oscillator response is obtained by inserting a microwave diode between the sample and the oscilloscope and is processed as the output signal.
The oscillation amplitude is robust to noise thanks to the confinement provided by the counteracting torques exerted by the injected current and the magnetic damping. In addition, the voltage amplitude is highly non-linear as a function of the injected current. The current depends on the voltage amplitude similarly to our simulated oscillator (square root dependence) in Eq. (\ref{Eq:test}). Furthermore, the amplitude of the oscillator voltage intrinsically depends on past inputs when the time step $\theta$ is shorter than the relaxation time of the magnetic nano-oscillator. Therefore, this single nano-device has the two most crucial properties of neurons: non-linearity and memory.

\begin{table}[ht]
\centering%
\def\arraystretch{1.5}%
\begin{tabular}{c|c|c|c|c|c}
\hline
   \multirow{2}{*}{Filter} & \multicolumn{2}{c|}{Training} & \multicolumn{2}{c|}{Testing} & Overfitting \\
   \cline{2-6}
                             & WSR (std) & MSE (std) & WSR (std) & MSE (std) & $\text{MSE}^\text{test}/\text{MSE}^\text{train}$ \\
   \hline
   Cochleagram               & 100.0~\% (0.0\%) & 0.0133 ($1.1\cdot10^{-4}$) & 99.6~\% (0.8~\%) & 0.0193 ($2.3\cdot10^{-3}$) & 1.451 \\
   \hline
   MFCC                      & 99.8~\% (0.1~\%) & 0.0259 ($1.1\cdot10^{-4}$) & 99.2~\% (1.0~\%) & 0.0283 ($8.2\cdot10^{-4}$) & 1.093 \\
   \hline
   $\Re(\text{Spectro})$ & 75.7~\% (0.6~\%) & 0.0708 ($1.1\cdot10^{-4}$) & 65.2~\% (4.2~\%) & 0.0738 ($7.3\cdot10^{-4}$) & 1.042 \\
   \hline
   Spectro HP & 99.2~\% (0.2~\%) & 0.0362 ($9.9\cdot10^{-5}$) & 98.6~\% (1.9~\%) & 0.0384 ($6.5\cdot10^{-4}$) & 1.061 \\
   \hline
\end{tabular}
\caption{\label{tab:tab1}Results for a simulated STNO neural network with $N = 400$ nodes.}
\end{table}

\begin{table}[ht]
\centering%
\def\arraystretch{1.5}%
\begin{tabular}{c|c|c|c|c|c}
\hline
   \multirow{2}{*}{Filter} & \multicolumn{2}{c|}{Training} & \multicolumn{2}{c|}{Testing} & Overfitting \\
   \cline{2-6}
                        & WSR (std) & MSE (std) & WSR (std) & MSE (std) & $\text{MSE}^\text{test}/\text{MSE}^\text{train}$ \\
   \hline
   Cochleagram          & 100.0~\% (0.0~\%) & 0.0192 ($1.1\cdot10^{-4}$) & 99.6~\% (0.8~\%) & 0.0222 ($1.3\cdot10^{-3}$) & 1.156 \\
   \hline
   MFCC                 & 99.8~\% (0.1~\%) & 0.0262 ($1.4\cdot10^{-4}$) & 99.8~\% (0.6~\%) & 0.0274 ($1.0\cdot10^{-3}$) & 1.046 \\
   \hline
   $\Re(\text{Spectro})$ & 90.2~\% (0.4~\%) & 0.0699 ($1.5\cdot10^{-4}$) & 80.0~\% (4.1~\%) & 0.0726 ($1.5\cdot10^{-3}$) & 1.039 \\
   \hline
   Spectro HP & 100.0~\% (0.0~\%) & 0.0327 ($7.1\cdot10^{-5}$) & 99.8~\% (0.6~\%) & 0.0344 ($7.7\cdot10^{-4}$) & 1.052 \\
   \hline
\end{tabular}
\caption{\label{tab:tab2}Results for a experimental STNO neural network with $N = 400$ nodes.}
\end{table}

Tables \ref{tab:tab1} and \ref{tab:tab2} show the word success rates, as well as mean squared error obtained by simulations and experimentally. The gain on the spoken digit recognition for the different acoustic filters induced by the experimental magnetic nano-oscillator is shown in Fig.~\ref{fig4}(b). There is very good agreement between the experimental results and the simulations. When 9 data sets are used while doing the training process the gain is 3.8~\% for the cochleagram, 10.8~\% for the {\SpectroHP} filter, 22~\% for the MFCC filter, and 70.4~\% for linear spectrogram. We see in Tables \ref{tab:tab1} and \ref{tab:tab2}, as well as Fig.~\ref{fig4} that, for some cases, the magnetic nano-oscillator exhibits slightly higher recognition gain than the simulations even though the latter neglects the intrinsic noise. We believe the better performance is mainly due to the higher complexity in the dynamics of the magnetic nano-oscillators, including a relaxation time that varies with current. Finally, the observation shows that overfitting effects are quite minimal. Some overfitting with respect to the mean squared error (MSE) can be seen when using cochleagram filters. This is due to the fact that in this situation, the role of the reservoir is quite minimal, whereas it uses the same number of parameters as in the other situations.  However, the overfitting occurs in the MSE does not make the overall performance for the word success rate (WSR) on the test set significantly worse.

\begin{figure}[ht]
    \centering
	\includegraphics[scale=0.7]{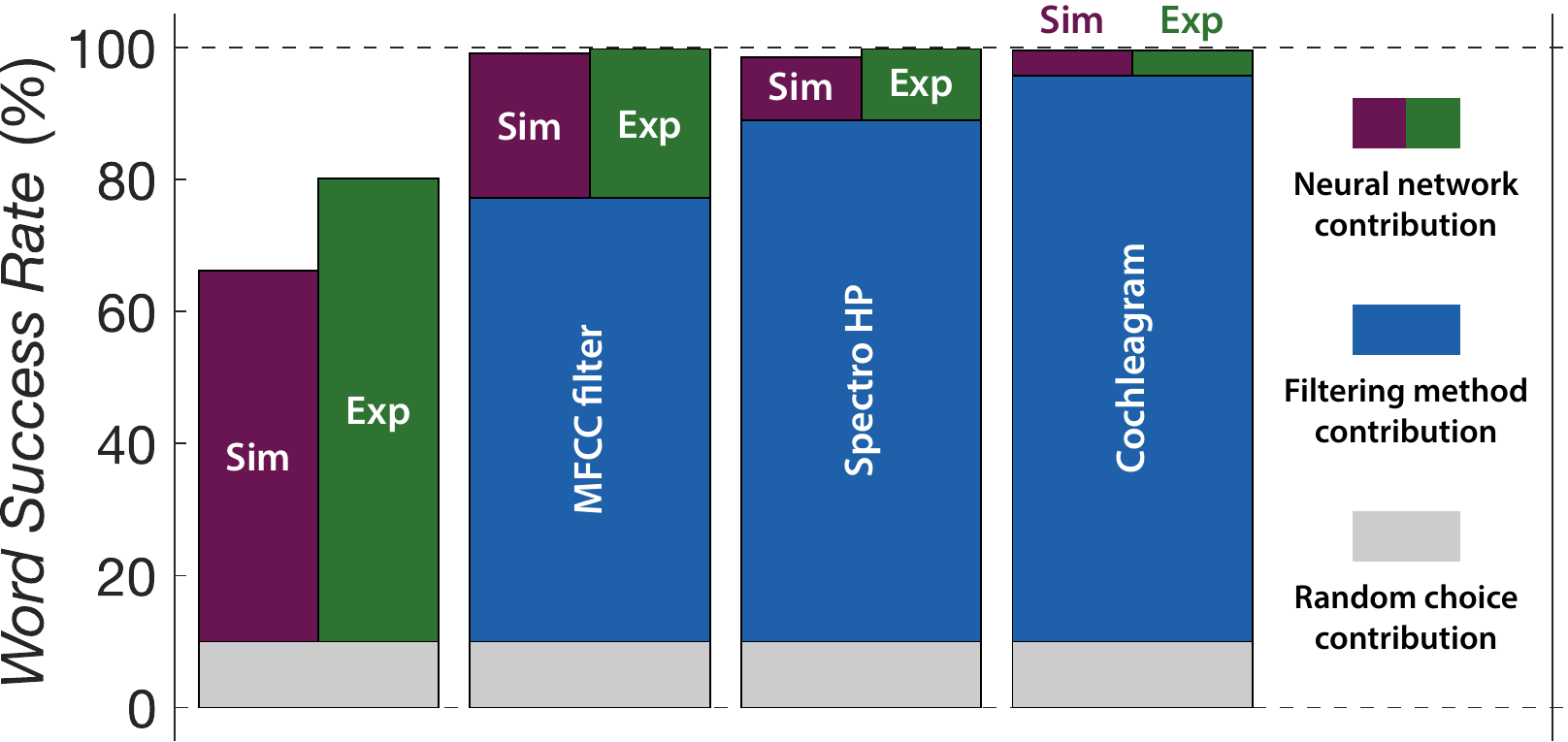}
	\caption{{\bf Contributions to the spoken digit cross-validated test recognition rate.} Random choice level is shown in grey, the filtering methods in blue, and the neural network under the reservoir computing approach in purple and green for the simulations and experiments, respectively. Here, 9 data subsets (90~\% of the database) are used for training our reservoir computing model and the remaining subset (10~\% of the database) is used to perform the recognition task.}
	\label{fig5}
\end{figure}

The different contributions to the spoken digit recognition task are summarised in Fig.~\ref{fig5} for the case in which nine utterances are used during the learning step and one during the recognition. The random choice level is 10~\% and is shown in grey. The contribution of the filtering methods is shown in blue (not visible in the case of the linear spectrogram, $\alpha = 1$). Fig.~\ref{fig5} also shows the net contribution of a neural network, in our case under the reservoir computing approach, to the spoken digit task. The simulated version of our neural network, i.e. using the simulated dynamics of the spin-torque vortex oscillators is shown in purple, while the results for the experimental magnetic nano-oscillators are shown in green. The main contribution to the spoken recognition task brought by the neural network happens when there is a lot of work to perform, i.e. when starting from the random choice level (linear spectrogram). Nevertheless, when our neural network is coupled with well performing stand-alone feature extraction techniques like the cochleagram or the MFCC filter, it is capable of bringing the recognition rate level to state-of-the-art values (overall WSR of 99.8~\% for the MFCC and the Spectro HP filters + experimental spin-torque vortex oscillator).

\section*{More challenging spoken digit database}

The TI-46 database is based on clean audio waveforms from a limited number of speakers.  We describe above how this database is rather limited for testing a neural network when combined with substantive preprocessing. We are able to test our nano-oscillator based reservoir computing approach by limiting the preprocessing to a basic linear preprocessing filter (linear spectrogram) and demonstrate its effectiveness. However, we can test the combination of effective preprocessing with reservoir computing by using a larger data set with a broader set of speakers and a variety of background noise types and levels. In this section, we simulate the performance of our theoretical reservoir computing implementation (simulated STNO neural network) on the AURORA-2 database \cite{Hirsch,ELRA}.

The AURORA-2 database provides the data for the task of recognising digits taken from the TIDigits database \cite{Leonard} in noise and channel distorted environments (artificially corrupted). To simulate noisy telephony environments, the clean utterances are first down-sampled to 8 kHz, and then additive and convolutional noise is added. The AURORA-2 database has both clean and multi-condition training and test sets. Each type of noise is added into a subset of clean speech utterances, with seven different levels of signal-to-noise ratios (SNRs). This process generates seven subgroups of test sets for a specified noise type, with clean (infinite signal-to-noise ratios) and signal-to-noise ratios of 20, 15, 10, 5, 0, and -5 dB.

We simulate the preprocessing and reservoir computing on all available isolated spoken digits from 0 to 9 in the training and testA datasets from 214 female and male speakers. The training dataset contains 2196 clean digits and several subsets of noisy (corrupted) digits with 4 different noise types (subway, babble, car, and exhibition hall noise) at different signal-to-noise ratios. During the training of our model, we select the 2196 clean digits and the corrupted digits with SNR = 20~dB (451 digits), SNR = 15~dB (444 digits), and SNR = 10~dB (430 digits). With this dataset, we train the model in mixed conditions.

The recognition (testing) is performed on testA dataset containing 4 types of added noise at different signal-to-noise ratios. The different noise types are the same as in the training set (unlike for the testB and testC subsets of AURORA-2 that we do not use). The testA dataset contains a subset of 1040 clean digits and several subsets containing each 1040 digits corrupted with subway, babble, car, and exhibition hall noise. We choose the following 4 subsets : clean and with corrupted with SNR = {20~dB, 15~dB, and 10~dB}.  To summarise, training is performed on 3521 digits and testing on 1040 unseen digits from the same  categories as the training set. The test set contains 22.8~\% of the total number of digits (1040/4561).

Tables \ref{tab:tab3} and \ref{tab:tab4} give the simulated results for spoken digit recognition using the nano-oscillator based reservoir computing approach combined with the two filtering methods, MFCC, and cochlear, respectively. The results are given in word success rate (\%). In parenthesis, we give the gain compared to the baseline (control test without the nano-oscillator based reservoir).  Not surprisingly, the results are not as good as the results presented above for the TI-46 database.  While the preprocessing filters do much worse without the reservoir than they do on the TI-46 database, they still do much better than linear preprocessing. In all cases, inclusion of the reservoir substantially improves the success rate compared to the baseline.

\begin{table}[ht]
\centering%
\def\arraystretch{1.5}%
\begin{tabular}{c|c|c|c|c|c}
\hline
SNR (dB) & Subway         & Babble         & Car            & Exhibition     & AVG \\
\hline
clean    & 95.15 (+56.34) & 94.96 (+54.65) & 90.98 (+45.88) & 90.73 (+45.94) & 92.96 (+50.70) \\
20       & 83.58 (+49.62) & 90.70 (+57.75) & 85.88 (+50.59) & 74.90 (+43.63) & 83.77 (+50.40) \\
15       & 79.48 (+47.39) & 85.66 (+54.65) & 83.92 (+51.76) & 68.73 (+44.79) & 79.45 (+49.65) \\
10       & 64.93 (+39.18) & 79.46 (+55.82) & 72.16 (+46.28) & 57.92 (+36.30) & 68.62 (+44.40) \\
\hline
AVG      & 80.79 (+48.13) & 87.70 (+55.72) & 83.24 (+48.63) & 73.07 (+42.67) & 81.20 (+48.79) \\
\hline
\end{tabular}
\caption{\label{tab:tab3}Word success rate (in percent) for a simulated STNO neural network with $N = 2000$ nodes after filtering the inputs with the MFCC filter combined with the reservoir.  These results are for a reservoir trained with all noise levels and types and then tested for each part of the test set with the different noise levels and types.  Each entry gives the overall word recognition rate and in parentheses, the gain achieved by adding the reservoir to the preprocessing filter.  The first column gives the noise level for different parts of the test set and the top row gives the noise type.  The bottom row gives the results over all levels of noise for each noise type and the right column gives the results for each noise level over all types of noise.  The bottom right entry is the average result over the whole test set.}
\end{table}

\begin{table}[ht]
\centering%
\def\arraystretch{1.5}%
\begin{tabular}{c|c|c|c|c|c}
\hline
SNR (dB) & Subway         & Babble         & Car            & Exhibition     & AVG \\
\hline
clean    & 88.81 (+23.88) & 87.21 (+24.81) & 89.02 (+29.02) & 91.51 (+25.87) & 89.14 (+25.90) \\
20       & 67.54 (+23.51) & 69.77 (+20.16) & 69.02 (+26.67) & 56.76 (+22.01) & 65.77 (+23.09) \\
15       & 70.15 (+27.61) & 63.57 (+16.96) & 68.63 (+23.14) & 59.46 (+28.57) & 65.45 (+24.07) \\
10       & 55.97 (+18.66) & 53.49 (+14.73) & 56.86 (+21.17) & 53.28 (+21.62) & 54.90 (+19.05) \\
\hline
AVG      & 70.62 (+23.42) & 68.51 (+19.17) & 70.88 (+25.00) & 65.25 (+24.52) & 68.82 (+23.02) \\
\hline
\end{tabular}
\caption{\label{tab:tab4} Word success rate (in percent) for a simulated STNO neural network with $N = 2000$ nodes after filtering the inputs with the cochlear filter combined with the reservoir.  These results are for a reservoir trained with all noise levels and types and then tested for each part of the test set with the different noise levels and types.  Each entry gives the overall word recognition rate and in parentheses, the gain achieved by adding the reservoir to the preprocessing filter.  The first column gives the noise level for different parts of the test set and the top row gives the noise type.  The bottom row gives the results over all levels of noise for each noise type and the right column gives the results for each noise level over all types of noise.  The bottom right entry is the average result over the whole test set.}
\end{table}

Comparison of Table \ref{tab:tab3} and Table \ref{tab:tab4} shows that the MFCC filter is more robust to noise than the cochlear filter and gives better results in most cases.  Interestingly, it does worse in almost all cases without the reservoir, but appears to allow the reservoir to make much larger improvements in the success rate. The average word success rate of the testing set containing only clean digits is 92.96~\% for the MFCC filter. The training was performed on clean and corrupted digits (mixed conditions). The improvement over the baseline is 50.70~\% (given in parentheses) implying that the baseline value is 42.26~\% (=  92.96~\% $-$ 50.70~\%). When the cochlear filter is used, the average gain brought by the neural network is +25.90~\% when testing clean digits and is about two times smaller than for the MFCC filter. The same holds when noisy conditions are tested. The overall gain is +48.79~\% (+23.02~\%) for an overall average recognition rate of 81.20~\% (68.82~\%) when the input is preprocessed with the MFCC filter (cochlear filter). The baseline is lower for the MFCC filter than for the cochlear filter but as the gain is much larger (about 2 times), the absolute performance of our neural network is larger in noisy conditions for the MFCC filter. We suspect that similar results would hold for the class of MFCC-like filters, some of which are even more robust against the inclusion of noise.

There are many differences between the simulations performed on the TI-46 and AURORA-2 databases. For the TI-46 database, there are only 5 female speakers uttering each digit 10 times. Training is performed on some utterances and recognition is performed on the others and the success rate is the average success rate over all combinations. For the AURORA-2 database, there are 214 speakers, half of them are female and half are male, and they typically utter each digit twice. In contrast to TI-46, the speakers in the training set are different from the speakers in the testing set. So, even without added noise, the test is much more difficult and involves almost 9 times more digits than for TI-46 (4561 digits in AURORA-2 vs 500 digits in TI-46).

Our results without the reservoir are consistent with previous results.  Cochlear filtering achieves approximately a 60~\% word success rate (around 40~\% for MFCC) by itself on clean isolated digits of AURORA-2 database.\cite{Larger2017} We obtain 63.24~\% (89.14~\% - 25.90~\%, see the last column of the first line of Table \ref{tab:tab4}) for the cochlear filter and 42.26~\% (92.96~\% - 50.70~\%, see the last column of the first line of Table \ref{tab:tab3}) for the MFCC filter.

\section*{Conclusion}

We test different frequency filtering methods as stand alone feature extractors. Training a linear classifier on the $\Rm$ vectors for the classic TI-46 spoken digit data base, both the cochleagram and the MFCC filter give high identification rates without further processing. On the other hand, the real part of a linear spectrogram does not separate the inputs of different digit classes.  Non-linearly transforming the spectrogram, gives similar results to the cochleagram and MFCC filters, stressing that the separation found for the MFCC and cochlear classifiers is due to the presence of non-linearity, with a minor effect due to the particular type of non-linear transformation.

In a second part, a non-linear oscillator is added to process the filtered input. The gain in word recognition due to the non-linear oscillator is computed for each filtering method. The non-linear oscillator is simulated and found to be in excellent agreement with experimental results with magnetic nano-oscillators. For the non-linear methods MFCC, \SpectroHP, and cochleagram, the gain of word recognition is small, despite a nearly perfect word recognition. On the the other hand, for  the linear spectrogram, the gain of word recognition is much higher even if the final word recognition is maximum 80$~\%$.

An important lesson is that when evaluating hardware systems with speech recognition tasks, the final word recognition rate should be interpreted with caution. If a very efficient filtering is used to preprocess the input, the hardware system may not be adding much performance. A hardware system only adds something if it provides improved word recognition. It should be noted that the use of more complicated datasets, such as the proprietary spoken digits dataset  used in \cite{Larger2017, Penkovsky2018}, or the inclusion of babble noise in the dataset would lead to significantly different results. The takeaway of our work is that, in order to test and compare hardware systems, using a linear spectrogram eases the interpretation of the results, because it does not introduce any separation of the input prior the hardware system. Furthermore, we show that a simple but powerful transformation like our {\SpectroHP} filter starting from a simple spectrogram achieves state-of-the-art results (simulations and experimentally) without applying any specific acoustic filter that mimics the human auditory system (like the cochleagram or the MFCC filter).

Testing on the AURORA-2 dataset reveals that under noisy conditions the cochlear filter performs better by itself than the MFCC filter but the gain brought by the neural network is two times better on average for the MFCC filter and the overall word success rate is higher for the MFCC filter than for the cochlear filter.

\bibliography{biblio}

\begin{thebibliography}{10}
\urlstyle{rm}
\expandafter\ifx\csname url\endcsname\relax
  \def\url#1{\texttt{#1}}\fi
\expandafter\ifx\csname urlprefix\endcsname\relax\def\urlprefix{URL }\fi
\expandafter\ifx\csname doiprefix\endcsname\relax\def\doiprefix{DOI: }\fi
\providecommand{\bibinfo}[2]{#2}
\providecommand{\eprint}[2][]{\url{#2}}

\bibitem{LeCun2015}
\bibinfo{author}{LeCun, Y.}, \bibinfo{author}{Bengio, Y.} \&
  \bibinfo{author}{Hinton, G.}
\newblock \bibinfo{journal}{\bibinfo{title}{Deep learning}}.
\newblock {\emph{\JournalTitle{Nature}}} \textbf{\bibinfo{volume}{521}},
  \bibinfo{pages}{436} (\bibinfo{year}{2015}).

\bibitem{Mead1990}
\bibinfo{author}{Mead, C.}
\newblock \bibinfo{journal}{\bibinfo{title}{Neuromorphic electronic systems}}.
\newblock {\emph{\JournalTitle{Proceedings of the IEEE}}}
  \textbf{\bibinfo{volume}{78}}, \bibinfo{pages}{1629--1636},
  \doiprefix\url{10.1109/5.58356} (\bibinfo{year}{1990}).

\bibitem{Mead2012}
\bibinfo{editor}{Mead, C.} \& \bibinfo{editor}{Ismail, M.} (eds.)
  \emph{\bibinfo{title}{Analog VLSI implementation of neural systems}},
  vol.~\bibinfo{volume}{80} (\bibinfo{publisher}{Springer Science \& Business
  Media}, \bibinfo{year}{2012}).

\bibitem{Maass2002}
\bibinfo{author}{Maass, W.}, \bibinfo{author}{Natschl{\"a}ger, T.} \&
  \bibinfo{author}{Markram, H.}
\newblock \bibinfo{journal}{\bibinfo{title}{Real-time computing without stable
  states: A new framework for neural computation based on perturbations}}.
\newblock {\emph{\JournalTitle{Neural Computation}}}
  \textbf{\bibinfo{volume}{14}}, \bibinfo{pages}{2531--2560},
  \doiprefix\url{10.1162/089976602760407955} (\bibinfo{year}{2002}).

\bibitem{Jaeger2004}
\bibinfo{author}{Jaeger, H.} \& \bibinfo{author}{Haas, H.}
\newblock \bibinfo{journal}{\bibinfo{title}{Harnessing nonlinearity: Predicting
  chaotic systems and saving energy in wireless communication}}.
\newblock {\emph{\JournalTitle{Science}}} \textbf{\bibinfo{volume}{304}},
  \bibinfo{pages}{78--80}, \doiprefix\url{10.1126/science.1091277}
  (\bibinfo{year}{2004}).

\bibitem{Appeltant2011}
\bibinfo{author}{Appeltant, L.} \emph{et~al.}
\newblock \bibinfo{journal}{\bibinfo{title}{Information processing using a
  single dynamical node as complex system}}.
\newblock {\emph{\JournalTitle{Nature Communications}}}
  \textbf{\bibinfo{volume}{2}}, \bibinfo{pages}{468} (\bibinfo{year}{2011}).

\bibitem{Paquot2012}
\bibinfo{author}{Paquot, Y.} \emph{et~al.}
\newblock \bibinfo{journal}{\bibinfo{title}{Optoelectronic reservoir
  computing}}.
\newblock {\emph{\JournalTitle{Scientific Reports}}}
  \textbf{\bibinfo{volume}{2}}, \bibinfo{pages}{287} (\bibinfo{year}{2012}).

\bibitem{Torrejon2017}
\bibinfo{author}{Torrejon, J.} \emph{et~al.}
\newblock \bibinfo{journal}{\bibinfo{title}{Neuromorphic computing with
  nanoscale spintronic oscillators}}.
\newblock {\emph{\JournalTitle{Nature}}} \textbf{\bibinfo{volume}{547}},
  \bibinfo{pages}{428} (\bibinfo{year}{2017}).

\bibitem{Davis1980}
\bibinfo{author}{Davis, S.} \& \bibinfo{author}{Mermelstein, P.}
\newblock \bibinfo{journal}{\bibinfo{title}{Comparison of parametric
  representations for monosyllabic word recognition in continuously spoken
  sentences}}.
\newblock {\emph{\JournalTitle{IEEE Transactions on Acoustics, Speech, and
  Signal Processing}}} \textbf{\bibinfo{volume}{28}},
  \bibinfo{pages}{357--366}, \doiprefix\url{10.1109/TASSP.1980.1163420}
  (\bibinfo{year}{1980}).

\bibitem{Verstraeten2005}
\bibinfo{author}{Verstraeten, D.}, \bibinfo{author}{Schrauwen, B.},
  \bibinfo{author}{Stroobandt, D.} \& \bibinfo{author}{Van~Campenhout, J.}
\newblock \bibinfo{journal}{\bibinfo{title}{Isolated word recognition with the
  liquid state machine: a case study}}.
\newblock {\emph{\JournalTitle{Information Processing Letters}}}
  \textbf{\bibinfo{volume}{95}}, \bibinfo{pages}{521 -- 528},
  \doiprefix\url{https://doi.org/10.1016/j.ipl.2005.05.019}
  (\bibinfo{year}{2005}).
\newblock \bibinfo{note}{Applications of Spiking Neural Networks}.

\bibitem{Hinton2012}
\bibinfo{author}{Hinton, G.} \emph{et~al.}
\newblock \bibinfo{journal}{\bibinfo{title}{Deep neural networks for acoustic
  modeling in speech recognition}}.
\newblock {\emph{\JournalTitle{IEEE Signal Processing Magazine}}}
  \textbf{\bibinfo{volume}{29}}, \bibinfo{pages}{82--97}
  (\bibinfo{year}{2012}).

\bibitem{Lyon1982}
\bibinfo{author}{Lyon, R.}
\newblock \bibinfo{title}{A computational model of filtering, detection, and
  compression in the cochlea}.
\newblock In \emph{\bibinfo{booktitle}{ICASSP '82. IEEE International
  Conference on Acoustics, Speech, and Signal Processing}},
  vol.~\bibinfo{volume}{7}, \bibinfo{pages}{1282--1285},
  \doiprefix\url{10.1109/ICASSP.1982.1171644} (\bibinfo{year}{1982}).

\bibitem{Slaney1988}
\bibinfo{author}{Slaney, M.}
\newblock \emph{\bibinfo{title}{Lyon's Cochlear Model}}.
\newblock \bibinfo{organization}{Apple Computer, Inc.},
  \bibinfo{address}{Cupertino, CA}, \bibinfo{edition}{apple computer technical
  report\#13} edn. (\bibinfo{year}{1988}).

\bibitem{Brunner2013}
\bibinfo{author}{Brunner, D.}, \bibinfo{author}{Soriano, M.~C.},
  \bibinfo{author}{Mirasso, C.~R.} \& \bibinfo{author}{Fischer, I.}
\newblock \bibinfo{journal}{\bibinfo{title}{Parallel photonic information
  processing at gigabyte per second data rates using transient states}}.
\newblock {\emph{\JournalTitle{Nature Communications}}}
  \textbf{\bibinfo{volume}{4}}, \bibinfo{pages}{1364} (\bibinfo{year}{2013}).

\bibitem{Vandoorne2014}
\bibinfo{author}{Vandoorne, K.} \emph{et~al.}
\newblock \bibinfo{journal}{\bibinfo{title}{Experimental demonstration of
  reservoir computing on a silicon photonics chip}}.
\newblock {\emph{\JournalTitle{Nature Communications}}}
  \textbf{\bibinfo{volume}{5}}, \bibinfo{pages}{3541} (\bibinfo{year}{2014}).

\bibitem{Riou2017_IEDM_Paper}
\bibinfo{author}{Riou, M.} \emph{et~al.}
\newblock \bibinfo{journal}{\bibinfo{title}{Neuromorphic computing through
  time-multiplexing with a spin-torque nano-oscillator}}.
\newblock {\emph{\JournalTitle{Electron Devices Meeting (IEDM), 2017 IEEE
  International}}} \bibinfo{pages}{36.3.1--36.3.4},
  \doiprefix\url{10.1109/IEDM.2017.8268505} (\bibinfo{year}{2017}).

\bibitem{Larger2012}
\bibinfo{author}{Larger, L.} \emph{et~al.}
\newblock \bibinfo{journal}{\bibinfo{title}{Photonic information processing
  beyond turing: an optoelectronic implementation of reservoir computing}}.
\newblock {\emph{\JournalTitle{Opt. Express}}} \textbf{\bibinfo{volume}{20}},
  \bibinfo{pages}{3241--3249}, \doiprefix\url{10.1364/OE.20.003241}
  (\bibinfo{year}{2012}).

\bibitem{Dejonckheere2014}
\bibinfo{author}{Dejonckheere, A.} \emph{et~al.}
\newblock \bibinfo{journal}{\bibinfo{title}{All-optical reservoir computer
  based on saturation of absorption}}.
\newblock {\emph{\JournalTitle{Opt. Express}}} \textbf{\bibinfo{volume}{22}},
  \bibinfo{pages}{10868--10881}, \doiprefix\url{10.1364/OE.22.010868}
  (\bibinfo{year}{2014}).

\bibitem{Vinckier2015}
\bibinfo{author}{Vinckier, Q.} \emph{et~al.}
\newblock \bibinfo{journal}{\bibinfo{title}{High-performance photonic reservoir
  computer based on a coherently driven passive cavity}}.
\newblock {\emph{\JournalTitle{Optica}}} \textbf{\bibinfo{volume}{2}},
  \bibinfo{pages}{438--446}, \doiprefix\url{10.1364/OPTICA.2.000438}
  (\bibinfo{year}{2015}).

\bibitem{Brunner2018}
\bibinfo{author}{Brunner, D.} \emph{et~al.}
\newblock \bibinfo{journal}{\bibinfo{title}{Tutorial: Photonic neural networks
  in delay systems}}.
\newblock {\emph{\JournalTitle{Journal of Applied Physics}}}
  \textbf{\bibinfo{volume}{124}}, \bibinfo{pages}{152004},
  \doiprefix\url{10.1063/1.5042342} (\bibinfo{year}{2018}).
\newblock \eprint{https://doi.org/10.1063/1.5042342}.

\bibitem{Penkovsky2018}
\bibinfo{author}{Penkovsky, B.}, \bibinfo{author}{Larger, L.} \&
  \bibinfo{author}{Brunner, D.}
\newblock \bibinfo{journal}{\bibinfo{title}{Efficient design of
  hardware-enabled reservoir computing in fpgas}}.
\newblock {\emph{\JournalTitle{Journal of Applied Physics}}}
  \textbf{\bibinfo{volume}{124}}, \bibinfo{pages}{162101},
  \doiprefix\url{10.1063/1.5039826} (\bibinfo{year}{2018}).
\newblock \eprint{https://doi.org/10.1063/1.5039826}.

\bibitem{Duport2012}
\bibinfo{author}{Duport, F.}, \bibinfo{author}{Schneider, B.},
  \bibinfo{author}{Smerieri, A.}, \bibinfo{author}{Haelterman, M.} \&
  \bibinfo{author}{Massar, S.}
\newblock \bibinfo{journal}{\bibinfo{title}{All-optical reservoir computing}}.
\newblock {\emph{\JournalTitle{Opt. Express}}} \textbf{\bibinfo{volume}{20}},
  \bibinfo{pages}{22783--22795}, \doiprefix\url{10.1364/OE.20.022783}
  (\bibinfo{year}{2012}).

\bibitem{Martinenghi2012}
\bibinfo{author}{Martinenghi, R.}, \bibinfo{author}{Rybalko, S.},
  \bibinfo{author}{Jacquot, M.}, \bibinfo{author}{Chembo, Y.~K.} \&
  \bibinfo{author}{Larger, L.}
\newblock \bibinfo{journal}{\bibinfo{title}{Photonic nonlinear transient
  computing with multiple-delay wavelength dynamics}}.
\newblock {\emph{\JournalTitle{Phys. Rev. Lett.}}}
  \textbf{\bibinfo{volume}{108}}, \bibinfo{pages}{244101},
  \doiprefix\url{10.1103/PhysRevLett.108.244101} (\bibinfo{year}{2012}).

\bibitem{Martinenghi2014}
\bibinfo{author}{Romain~Martinenghi, M. J. Y. K. C. L.~L., Antonio
  Baylon~Fuentes}.
\newblock \bibinfo{journal}{\bibinfo{title}{Towards optoelectronic
  architectures for integrated neuromorphic computers}}.
\newblock {\emph{\JournalTitle{Proc. SPIE OPTO}}}
  \textbf{\bibinfo{volume}{8989}}, \doiprefix\url{10.1117/12.2038347}
  (\bibinfo{year}{2014}).

\bibitem{Larger2017}
\bibinfo{author}{Larger, L.} \emph{et~al.}
\newblock \bibinfo{journal}{\bibinfo{title}{High-speed photonic reservoir
  computing using a time-delay-based architecture: Million words per second
  classification}}.
\newblock {\emph{\JournalTitle{Phys. Rev. X}}} \textbf{\bibinfo{volume}{7}},
  \bibinfo{pages}{011015}, \doiprefix\url{10.1103/PhysRevX.7.011015}
  (\bibinfo{year}{2017}).

\bibitem{Lukosevicius2009}
\bibinfo{author}{Luko{\v s}evi{\v c}ius, M.} \& \bibinfo{author}{Jaeger, H.}
\newblock \bibinfo{journal}{\bibinfo{title}{Reservoir computing approaches to
  recurrent neural network training}}.
\newblock {\emph{\JournalTitle{Computer Science Review}}}
  \textbf{\bibinfo{volume}{3}}, \bibinfo{pages}{127 -- 149},
  \doiprefix\url{https://doi.org/10.1016/j.cosrev.2009.03.005}
  (\bibinfo{year}{2009}).

\bibitem{Maaten2008}
\bibinfo{author}{van~der Maaten, L.} \& \bibinfo{author}{Hinton, G.}
\newblock \bibinfo{journal}{\bibinfo{title}{Visualizing data using t-sne}}.
\newblock {\emph{\JournalTitle{Journal of machine learning research}}}
  \textbf{\bibinfo{volume}{9}}, \bibinfo{pages}{2579--2605}
  (\bibinfo{year}{2008}).

\bibitem{Slavin2009}
\bibinfo{author}{Slavin, A.~N.} \& \bibinfo{author}{Tiberkevich, V.~S.}
\newblock \bibinfo{journal}{\bibinfo{title}{Nonlinear auto-oscillator theory of
  microwave generation by spin-polarized current}}.
\newblock {\emph{\JournalTitle{IEEE Transactions on Magnetics}}}
  \textbf{\bibinfo{volume}{45}}, \bibinfo{pages}{1875 --1918},
  \doiprefix\url{10.1109/TMAG.2008.2009935} (\bibinfo{year}{2009}).

\bibitem{Grimaldi2014}
\bibinfo{author}{Grimaldi, E.} \emph{et~al.}
\newblock \bibinfo{journal}{\bibinfo{title}{Response to noise of a vortex based
  spin transfer nano-oscillator}}.
\newblock {\emph{\JournalTitle{Physical Review B}}}
  \textbf{\bibinfo{volume}{89}}, \bibinfo{pages}{104404--},
  \doiprefix\url{10.1103/PhysRevB.89.104404} (\bibinfo{year}{2014}).

\bibitem{Romera2018}
\bibinfo{author}{Romera, M.} \emph{et~al.}
\newblock \bibinfo{journal}{\bibinfo{title}{Vowel recognition with four coupled
  spin-torque nano-oscillators}}.
\newblock {\emph{\JournalTitle{Nature}}} \textbf{\bibinfo{volume}{563}},
  \bibinfo{pages}{230--234}, \doiprefix\url{10.1038/s41586-018-0632-y}
  (\bibinfo{year}{2018}).

\bibitem{Tsunegi2014}
\bibinfo{author}{Tsunegi, S.} \emph{et~al.}
\newblock \bibinfo{journal}{\bibinfo{title}{High emission power and q factor in
  spin torque vortex oscillator consisting of feb free layer}}.
\newblock {\emph{\JournalTitle{Applied Physics Express}}}
  \textbf{\bibinfo{volume}{7}}, \bibinfo{pages}{063009} (\bibinfo{year}{2014}).

\bibitem{Hirsch}
\bibinfo{author}{Hirsch, H.} \& \bibinfo{author}{Pearce, D.}
\newblock \bibinfo{title}{The aurora experimental framework for the performance
  evaluation of speech recognition systems under noisy conditions}.
\newblock In \emph{\bibinfo{booktitle}{Proceedings of ISCA ITRW ASR2000 on
  Automatic Speech recognition: Challenges for the next Millennium}}
  (\bibinfo{organization}{Paris, France}, \bibinfo{year}{2000}).

\bibitem{ELRA}
\emph{\bibinfo{title}{{ELRA} catalogue (http://catalog.elra.info), {AURORA}
  {P}roject {D}atabase, v2.0, {ISLRN}: 977-457-139-304-2, {ELRA} {ID}:
  {AURORA}/{CD}0002}}.

\bibitem{Leonard}
\bibinfo{author}{Leonard, R.} \& \bibinfo{author}{Doddington, G.}
\newblock \emph{\bibinfo{title}{{TID}igits}}.
\newblock \bibinfo{organization}{Linguistic Data Consortium},
  \bibinfo{address}{Philadelphia} (\bibinfo{year}{1993}).

\end{thebibliography}
%\noindent LaTeX formats citations and references automatically using the bibliography records in your .bib file, which you can edit via the project menu. Use the cite command for an inline citation, e.g.  \cite{Hao:gidmaps:2014}.
%For data citations of datasets uploaded to e.g. \emph{figshare}, please use the \verb|howpublished| option in the bib entry to specify the platform and the link, as in the \verb|Hao:gidmaps:2014| example in the sample bibliography file.

\section*{Acknowledgements}
%Acknowledgements should be brief, and should not include thanks to anonymous referees and editors, or effusive comments. Grant or contribution numbers may be acknowledged.

F.A.A. is a Research Fellow of the F.R.S.-FNRS. This work was supported by the European Research Council ERC under Grant bioSPINspired 682955.

\section*{Author contributions statement}

% Must include all authors, identified by initials, for example:
% A.A. conceived the experiment(s),  A.A. and B.A. conducted the experiment(s), C.A. and D.A. analysed the results.  All authors reviewed the manuscript.

The study was designed by F.A.A, J.G., and M.D.S., samples were optimised and fabricated by S.T. and K.Y., experiments were performed by M.R. and J.T., and numerical studies were realised by F.A.A.. F.A.A. wrote the core of the manuscript and all authors contributed to the text as well as to the analysis of the results.

\section*{Data Availability}

The datasets generated during and/or analysed during the current study are available from the corresponding author on reasonable request.

\section*{Additional information}

The authors declare no competing interests.

% To include, in this order: \textbf{Accession codes} (where applicable); \textbf{Competing interests} (mandatory statement). 

% The corresponding author is responsible for submitting a \href{http://www.nature.com/srep/policies/index.html#competing}{competing interests statement} on behalf of all authors of the paper. This statement must be included in the submitted article file.

\end{document}